\documentclass{aa}  
\usepackage{graphicx}
\usepackage{natbib}
\usepackage[breaklinks=true]{hyperref}
\usepackage{orcidlink}

\definecolor{amethyst}{rgb}{0.8, 0.0, 0.0}

\definecolor{orange}{rgb}{1,0.5,0}

\definecolor{boh}{rgb}{1,0,0}

\definecolor{vale}{RGB}{1,0,0}

\definecolor{giuliano}{RGB}{102,0,204}

\definecolor{laura}{RGB}{170, 100, 255}

\definecolor{green}{rgb}{0., 0.75, 0.}
\newcommand{\rev}[1]{{#1}}
\newcommand{\revv}[1]{{#1}}

\newcommand{\reved}[1]{{#1}}

\def\mum{{\mu \rm{m}}}

\def\Lduecinquanta{\nu L_\nu \, (250 \, \mum)}
\def\duecinquanta{{250\,\mum}}

\usepackage{txfonts}

\begin{document} 

   \title{Dust sub-millimetre emission in green valley galaxies}

   \author{Massimiliano Parente \inst{1,2,3}\orcidlink{0000-0002-9729-3721} 
    \and 
    Cinthia Ragone-Figueroa \inst{4,2,3}\orcidlink{0000-0003-2826-4799}
    \and
    Gian Luigi Granato \inst{2,3,4}\orcidlink{0000-0002-4480-6909}
    \and
    Laura Silva \inst{2,3}
    \and \\
    Valeria Coenda \inst{4,5} \orcidlink{0000-0001-5262-3822}
    \and
    Héctor J. Martínez \inst{4,5} \orcidlink{0000-0003-0477-5412}
    \and 
    Hernán Muriel \inst{4,5} \orcidlink{0000-0002-7305-9500}
    \and
    Andrea Lapi \inst{1,3,6,7} \orcidlink{0000-0002-4882-1735}
    }

    \institute
   {
    SISSA, Via Bonomea 265, I-34136 Trieste, Italy
    \and
    INAF, Osservatorio Astronomico di Trieste, via Tiepolo 11, I-34131, Trieste, Italy
    \and
    IFPU, Institute for Fundamental Physics of the Universe, Via Beirut 2, 34014 Trieste, Italy
    \and   
    IATE - Instituto de Astronom\'ia Te\'orica y Experimental, Consejo Nacional de Investigaciones Cient\'ificas y T\'ecnicas de la\\ Rep\'ublica Argentina (CONICET), Universidad Nacional de C\'ordoba, Laprida 854, X5000BGR, C\'ordoba, Argentina
    \and
    Observatorio Astronómico, Universidad Nacional de Córdoba, Laprida 854, X5000BGR, Córdoba, Argentina
    \and
   INFN, Instituto Nazionale di Fisica Nucleare, Via Valerio 2, I-34127, Trieste, Italy
    \and
   INAF/IRA, Istituto di Radioastronomia, Via Piero Gobetti 101, 40129 Bologna, Italy
\\
\\
    \email{mparente@sissa.it -- massimiliano.parente@inaf.it}
   }

    \date{Received XXX; Accepted YYY}

  \abstract
   {Green valley (GV) galaxies are objects defined on a colour--magnitude diagram, or a colour--mass diagram,  as being associated with a transition from a star-forming to a quiescent state (quenching), or vice versa (rejuvenation).}
   {We studied the sub-millimetre emission of galaxies in the GV and linked it with their physical evolutionary properties.}
   {We exploited a semi-analytic model (SAM) for galaxy evolution that includes a detailed treatment of dust production and evolution in galactic contexts. We modelled the observational properties of simulated galaxies by post-processing the SAM catalogues with the spectral synthesis and radiative transfer code GRASIL.}
   {Our model produces a clear bimodality (and thus a GV) in the colour--mass diagram, although some tensions arise when compared to observations.
   After introducing a new criterion for identifying the GV in any dataset, \rev{we find that GV galaxies, at fixed stellar mass, have $250 \mum$ luminosities approximately half those of blue galaxies, while red galaxies exhibit luminosities of up to an order of magnitude lower.} While specific star formation rates drop sharply during quenching, the dust content remains relatively high during the GV transition, powering sub-millimetre emission. Rejuvenating galaxies in the GV, which were previously red, have experienced a star formation burst that shifts their colour to green, but their $S_{250\, \mu \rm m}$ fluxes remain low due to their still low dust masses.}
   {Our galaxy evolution model highlights the delay between star formation and dust evolution, showing that sub-millimetre emission is not always a safe indicator of star formation activity, with quenching (rejuvenating) GV galaxies featuring relatively high (low) sub-millimetre emission.}

   \maketitle

\section{Introduction}

The `green valley' (GV; \citealt{Wyder:2007}) in galaxy evolution refers to the transitional region on a colour--magnitude (or colour--mass) diagram that lies between the blue cloud (BC) of star-forming galaxies and the red sequence (RS) of passive galaxies. This valley is believed to result from the relatively rapid transition of galaxies from the star-forming state to a passive one, making it crucial for understanding galaxy quenching processes (see \citealt{Salim2014} for a review). However, some inactive galaxies can experience `rejuvenation', a phase during which they undergo a new burst of star formation after a period of minimal activity. This rejuvenation can move galaxies in the RS back into the GV, and in some cases even into the BC (e.g. \citealt{Graham:2017}; \citealt{Rowlands2018}; \citealt{Chauke2019}).

Defining the GV is a complex task. Traditional approaches often rely on colour to define the GV \citep[e.g.][]{Brammer09,Schawinski2014,Gu2018}, but this can lead to systematic errors due to dust attenuation. \citet{Salim2014} highlights the use of the UV-optical colour--magnitude diagram for identifying the GV, which effectively distinguishes between star-forming and quiescent galaxies without the biases of optical colours. An alternative method proposed by \citet{Angthopo2019} employs the 4000 Å break strength, offering a dust-insensitive metric. \citet{Nyiransengiyumva2021} emphasise that different GV selection criteria can lead to variations in the resulting galaxy populations in terms of colour, specific star formation rate (sSFR), or star formation rate (SFR), demonstrating the challenges in defining the GV. \revv{Using the $NUVrK$ diagram, \cite{Gael22} introduced the parameter $\Delta_{\rm GV}$, which quantifies the distance from the lower boundary of the GV and indicates the likelihood of an object being in transition.} More recently, \citet{Pandey2024} introduced a novel definition using entropic thresholding on the colour-stellar mass plane, which provides a natural boundary for the GV. This wide array of methods and data, ranging from spectroscopic analyses of the Sloan Digital Sky Survey (SDSS) to UV observations from \reved{the Galaxy Evolution Explorer (GALEX)}, underscores both the complexity and importance of accurately defining the GV for studies of galaxy evolution.

Despite the various methods used to identify GV galaxies, they all share the common goal of identifying galaxies transitioning from a star-forming state to a quiescent, passive state. The properties of these GV galaxies can provide insights into the mechanisms driving quenching, such as halo heating \citep{Marasco2012}, supernova-driven winds (e.g. \citealt{Bower:2012, Stringer:2012}), feedback from massive stars (e.g. \citealt{DallaV:2008, Hopkins:2012}), and active galactic nucleus (AGN) feedback (e.g. \citealt{Granato2004, DiMatteo2005, Nandra:2007, Hasinger:2008, Silverman:2008, Cimatti:2013}). It has often been pointed out that AGN feedback, in particular, plays a relevant role due to the higher prevalence of AGNs in GV galaxies compared to those in the blue or red sequences (\citealt{Gu2018, Schawinski2010, Wang2017, Lacerda2020}; but see also \citealt{Mahoro2022}). Additionally, external quenching mechanisms, such as ram pressure and strangulation, can also drive the transition in the colour--magnitude diagram, making the environment in which galaxies reside another crucial factor in shaping the GV transition \citep[e.g.][]{Lin:2019, Bluck:2020,Trussler:2020, Sampaio:2022}. Numerous studies have shown that denser environments and galaxy groups facilitate more efficient quenching \citep{Coenda2018,GonzalezDelgado2022,Chang2022}.

Numerical simulations have been widely used to study the transition of galaxies within the GV. The insights from these theoretical models, such as NIHAO, IllustrisTNG, and EAGLE, indicate that AGN feedback is the primary mechanism driving the GV transition 
\citep[e.g.][]{Blank2022,Angthopo2021}. These simulations also offer valuable information on the timescales required for this transition, often referred to as quenching timescales, and how they relate to other galaxy properties, such as stellar mass and the surrounding environment \citep[e.g.][]{Nelson2018, Wright2019}.\\

In this work we focus on the sub-millimetre dust emission of galaxies within the GV. We explored this characteristic using a galaxy catalogue generated from a semi-analytic model (SAM) that includes a self-consistent model of dust formation and evolution. By applying a radiative transfer (RT) approach, we derived observable properties that account for both stellar emission and the absorption and re-emission by dust grains. 
After defining GV galaxies with a new criterion applicable to any dataset, \rev{we find that GV galaxies, at fixed stellar mass, have $250 \mum$ luminosities approximately half those of blue galaxies, while red galaxies exhibit luminosities of up to an order of magnitude lower.} This is because, despite a decline in the sSFR and the resulting green optical colours, the dust content remains high during the GV phase, fuelling the sub-millimetre emission. The differences in the evolution of the sSFR and dust content make sub-millimetre emission a potential indicator of rejuvenating galaxies, which transition to bluer optical colours while still exhibiting low sub-millimetre fluxes.

We introduce our simulated galaxy catalogue in Sect. \ref{sec:simucata}, and in particular the pipeline to obtain observable properties from it (Sect. \ref{sec:grasil}), while the observational data we compare our catalogue with are presented in Sect. \ref{sec:data}. After inspecting the colour--mass diagram predicted by our simulation in Sect. \ref{sec:cmd}, we present our new definition of the GV in Sect. \ref{sec:GVdefinition}, which is applied to both the simulated and observed sample of galaxies. We then analyse dust attenuation in GV galaxies in Sect. \ref{sec:dustatt}, while Sects. \ref{sec:250em} and \ref{sec:250em_evo} focus on dust $250\,\mum$ emission. The emission in systems undergoing rejuvenation is discussed in Sect. \ref{sec:rejuv}. Finally, we summarise our results and highlight our main conclusions in Sect. \ref{sec:conclusions}.

\section{Simulated galaxy catalogue}
\label{sec:simucata}

\subsection{The SAM}
\label{sec:sam}

We built our catalogue of simulated galaxies starting from the \textsc{L-Galaxies} SAM. In particular, we adopted its last public release \footnote{The source code is available at \url{https://github.com/LGalaxiesPublicRelease/LGalaxies_PublicRepository/releases/tag/Henriques2020}} \citep{Henriques2020}, along with the updates introduced and discussed in \cite{Parente2023}. Among these updates, of relevance to this work is the inclusion of a detailed model for the production and evolution of dust grains, inspired to the dust model adopted by our group in hydrodynamic simulations \citep{Granato21,Parente22}. The dust model (see Sect. 2.1 of \citealt{Parente2023} for an exhaustive description) includes two sizes and two chemical compositions of grains. Grains are produced in asymptotic giant branch stars envelopes and type-II supernovae ejecta and, thus, ejected into the surrounding gaseous medium. Different processes are then taken into account, which affect the mass (grains accretion, destruction in supernova shocks, and thermal sputtering) and the size (shattering and coagulation) evolution of grains. These processes depend on the physical properties of the galaxy as provided by the SAM (e.g. molecular gas fraction, hot gas temperature, and metallicity). The dust content and properties as predicted by the SAM are a crucial ingredient for predicting the observability of simulated galaxies, as detailed in Sect. \ref{sec:grasil}.\\ 

The SAM is run on top of the \textsc{Millennium} merger trees (\citealt{Springel05}, box size $500\ {\rm Mpc}/h$, $2160^3$ particles), and a Planck cosmology\footnote{The cosmology originally adopted in the \textsc{Millennium} simulation has been scaled according to \cite{Angulo2010} and \cite{Angulo2015}.} \citep{Planck14} is assumed throughout this work ($h=0.673$, $\Omega_{\rm m}= 0.315$, $\Omega_{\rm b}= 0.0487$, $\sigma_8=0.829$). A \cite{Chabrier2003} initial mass function is adopted, with with mass limits $0.1-100$ $M_\odot$. We analysed galaxies with a stellar mass content of at least $\log(M_{\rm stars}/M_\odot) \geq 9$, which approximately corresponds to the resolution limit of the underlying dark matter simulation \citep[e.g.][]{Guo11}.
In this work, to keep the analysis within a reasonable computational time, a reduced sample -- that uses the same \reved{Stellar Mass Function (SMF)} as the full simulation -- is post-processed to derive observable properties.

\subsection{Radiative transfer with GRASIL}
\label{sec:grasil}
The observable properties of our simulated galaxies are derived using the RT code GRASIL \citep{Silva98,Granato00}. This code computes the emission from stellar populations and the absorption and thermal emission from dust grains. It utilises the information on star formation histories (SFHs), the geometry of stars and ISM, and the abundance and properties of dust grains for each galaxy as predicted by the SAM, as detailed in the following sections. We refer to the above-mentioned papers for a detailed description of GRASIL.\\

\subsubsection{Stars and ISM geometry}

Simulated galaxies are assumed to have a disc-like interstellar medium (ISM), while the stellar component is organised into a disc and a bulge. Both the stellar and ISM discs follow a radial and vertical exponential profile, meaning the density of the disc is given by
\begin{equation}
    \rho \propto {\rm exp}\left(- r/r_{\rm d}\right) {\rm exp}\left(- \left| z \right|/z_{\rm d}\right), 
\end{equation}where $r$ and $z$ are the radial and vertical coordinates, and $r_{\rm d}$ and $z_{\rm d}$ the respective scale lengths of the exponential profiles. The scale length radii of the stellar and ISM discs are predictions from the SAM (note that these values usually differ). For both cases, we set the scale height to $0.1$ of the scale radius. The stellar bulge is assumed to have a King profile, with its core radius derived from the half-mass radius predicted by the SAM.

\subsubsection{Stellar emission}

The stellar emission of galaxies originates from the populations in the disc and bulge. Both components have an associated SFH\footnote{The same SFH is adopted within the SAM to compute the chemical enrichment from stars. However, although the SFH in the code is recorded on a ring-by-ring basis, for GRASIL post-processing we used the SFH integrated over the entire stellar disc and bulge.}, which tracks the time assembly of their stellar mass and metallicity. The SFH is stored in $N_{\rm SFH}$ time bins, with the size of each bin varying with time so that the most recent bins have the highest resolution (see Fig. S2 of the \citealt{Henriques2020} documentation for a schematic representation). The emission from stellar populations at time $t$ is calculated (for both disc and bulge stars) by summing over the SFH bins:

\begin{equation}
    L(t) = \sum _{i=0}^{N_{\rm SFH}} l_{i}(t-t_{i}; Z_i) \cdot \psi_i \Delta t_{i},
\end{equation}
where $l_{i}(t-t_{i}; Z_i)$ is the emission associated with a unit mass simple stellar population featuring age\footnote{Here $t_i$ is the central time of the $i-$th bin.} $t-t_{i}$ and metallicity $Z_i$, $\psi_i$, and $\Delta t_{i}$ are the SFR and time width associated with the $i-$th bin.\\

\subsubsection{Radiative transfer}

Radiation emitted from stars propagates into the ISM, where it is extinguished, heats dust grains, and causes their subsequent thermal emission. Modelling these processes requires detailed RT computations, which are performed by GRASIL.

The interaction between starlight and dust grains depends on the abundance and properties of the latter. The mass of dust in each galaxy, as well as the chemical composition and size of the grains, are predictions of the SAM. Specifically, the model predicts the abundance of four dust species, following two chemical compositions (silicate and carbonaceous) and two sizes with representative radii of $0.05 \, \mu {\rm m}$ and $0.005 \, \mu {\rm m}$ (referred to as large and small, respectively). GRASIL uses this information and considers a population of silicate and carbonaceous grains with two different grain size distributions, which are power laws shaped by the ratio of small to large grains. These information are needed to model the absorption and emission of grains.

Dust is assumed to be organised into two phases of the ISM: the dense molecular clouds (MCs) and the diffuse medium. Dust properties are assumed to be the same in the two regions. The fraction of the total dust mass residing in MCs is $f_{\rm mol}$, which is the fraction of the total ISM gas in molecular form. This parameter is predicted by the SAM, which relies on the \cite{Krumholz09} prescription.

MCs are assumed to have a mass $M_{\rm c}$ and a radius $r_{\rm c}$, whose combination $M_{\rm c}/r^2_{\rm c}$ is a parameter of GRASIL. This represents the optical depth of the clouds. Stars are assumed to form in these clouds and progressively escape according to an escape timescale, $t_{\rm esc}$. Specifically, the fraction of stars inside MCs decreases with time according to

\begin{equation}
    f(t) =
    \begin{cases}
    1 \qquad \qquad \qquad \qquad t \leq t_{\rm esc}, \\
    2 - {t}/{t_{\rm esc}} \quad \qquad \qquad t_{\rm esc} < t \leq 2t_{\rm esc}, \\
    0 \qquad \qquad \qquad \qquad t>2t_{\rm esc}. \\
    \end{cases}
    \label{eq:t_escape}
\end{equation}The escape time, $t_{\rm esc}$, is a GRASIL parameter and is set to $3\,{\rm Myr}$ here. Starlight propagation in MCs is simulated using full RT computations \citep{Granato94} because the high densities of these objects make the self-absorption of IR photons significant.
In contrast, starlight propagation in the diffuse medium (where IR photon self-absorption is negligible) is treated in a simplified manner by introducing an effective absorption optical depth, defined as $\tau_{\rm abs, , eff} = \sqrt{\tau_{\rm abs} \left( \tau_{\rm abs} + \tau_{\rm sca} \right)}$, where $\tau_{\rm abs}$ and $\tau_{\rm sca}$ are the true absorption and scattering optical depths, respectively.\\

The outcome of the pipeline described above is that, at each point of the galaxy, stellar radiation is absorbed, the temperature of the grains is calculated (with large grains reaching an equilibrium temperature and small grains experiencing stochastic temperature fluctuations), and the dust emission is modelled as a grey body. This results in a spectral energy distribution (SED) from $0.1\,\mu{\rm m}$ to $1000\,\mu{\rm m}$ for each galaxy along a given line of sight. The apparent magnitude of the galaxy in a specific band is then obtained by convolving the resulting SED with the appropriate filter.

\section{Observational data}
\label{sec:data}

The observational analysis presented in this work relies on data from the Galaxy And Mass Assembly (GAMA) survey\footnote{\url{https://www.gama-survey.org/}} \citep{Driver2009,Driver2011,Liske2015,Baldry2018,Driver2022}. This is a spectroscopic, large-scale survey covering a total area of $\sim 280 \, {\rm deg}^2$ across five different regions of the sky. Together with spectroscopic redshifts, this survey provides pancromatic photometry for $\gtrsim 200,000$ galaxies obtained combining data from other surveys operating from the X-ray to the FIR side of the spectrum. Moreover, many other data products obtained from SED fitting are available, such as stellar masses, SFRs, and dust masses.

We employed the last GAMA data release, DR4 \citep{Driver2022}. In particular, we used GAMA II\footnote{The second GAMA surveying phase.} data for the three equatorial survey regions G09, G12, and G15, which all have a limiting Petrosian $r-$band apparent magnitude of $19.8$. We selected galaxies with high quality redshift measurements in the range $0.002 < z <0.08$.
In our analysis we are interested in the optical colour $u-r$ and the sub-millimetre $250\,\mum$ emission of galaxies. GAMA provides both quantities, the former coming from the SDSS (\citealt{Abazajian2009}) and the latter from the \reved{Herschel Astrophysical Terahertz Large Area Survey (\textit{H}-ATLAS)} survey \citep{Eales2010,Valiante2016}. Here we used the matched aperture photometry fluxes obtained from the \textsc{LambdaR} code \citep{Wright2016} and selected galaxies with a signal-to-noise ratio $>3$. Fluxes are corrected for galactic extinction and $k-$corrected to $z=0$ (i.e. they are rest-frame; \citealt{Loveday2012}). Finally, we used stellar masses obtained from this photometry by SED fitting \citep{Taylor2011}, selecting galaxies with $M_{\rm stars}\geq 10^9 \, M_\odot$, as in our simulated sample. 
\rev{Since the original catalogue is flux-complete down to a apparent magnitude of $r=19.8$, we applied $1/V_{\rm max}$ weights \citep{Schmidt:1968A} to all
GAMA galaxies in our \revv{analysis of the colour-mass diagram (Sect. \ref{sec:cmd})}. We note that these weights are equal to unity for all but for a very minor fraction of galaxies with $M_{\rm stars}\lesssim 10^{9.3} M_{\odot}$.}

\section{The colour--mass diagram}
\label{sec:cmd}

\begin{figure}
\centering
\includegraphics[width=1\hsize]{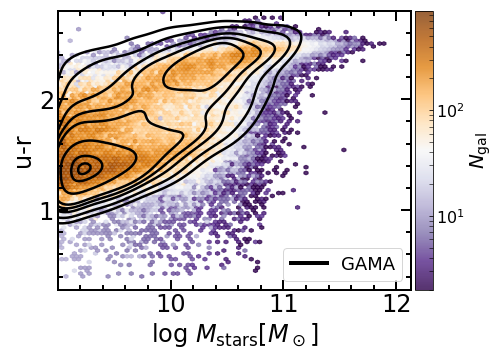}
  \caption{Colour $(u-r)-$stellar mass diagram at $z \leq 0.08$ for our simulated galaxies (2D histogram). Contour lines are $1,\,5,\,20,\,50,\, 70,\,80$, and $\,90\,\%$ of the distribution and refer to the same colour--mass relation in the GAMA sample adopted in this work. The $u-r$ colour is dust-attenuated and $k$-corrected to $z=0$.}
     \label{fig:colorMass}
\end{figure}

Before analysing any properties of GV galaxies in our catalogues, it is worth looking at the distribution of simulated galaxies in the colour--mass diagram.
The colour $(u-r)$ $-$ stellar mass diagram for local galaxies ($z \leq 0.08$) is reported in Fig. \ref{fig:colorMass}. For comparison, we also report observations from the GAMA sample as black contours. The reported colours are $k$-corrected to $z=0$ and dust-attenuated, meaning that they are not intrinsic stellar colours. 

Our model produces a clear population of blue galaxies at $M_{\rm stars}\lesssim 10^{10.5}\, M_\odot$, which matches the observed BC  well. As for red galaxies, in our model these are typically more massive than the blue ones, in line with observations. However, our model overproduces the abundance of low mass ($M_{\rm stars}\lesssim 10^{10}\,M_\odot$) red galaxies. These are mainly satellite systems, and in particular orphan galaxies. This has been noted recently by \cite{Harrold2024} in the context of \textsc{L-Galaxies}. These authors highlighted the overabundance of orphans in the low mass SMF at $0.5<z<3$, and claimed that increasing star formation in these systems could help resolve this problem.
A similar distribution, and similar issues as well, is observed when adopting near-UV$-r$ or $g-r$ colours, not shown here for brevity.\\

\section{Defining the GV}
\label{sec:GVdefinition}

\begin{figure*}[htb]
    \centering
    \includegraphics[width=0.6\columnwidth]{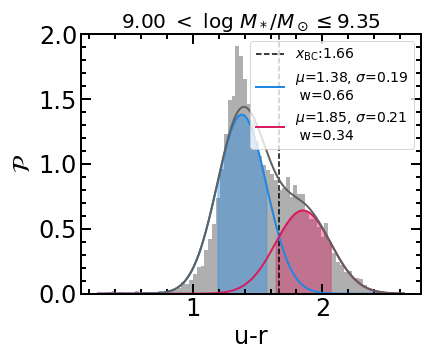}\quad
    \includegraphics[width=0.6\columnwidth]{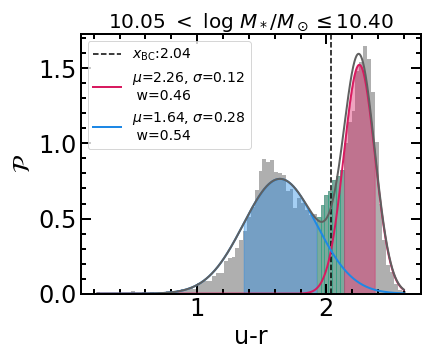}\quad
    \includegraphics[width=0.6\columnwidth]{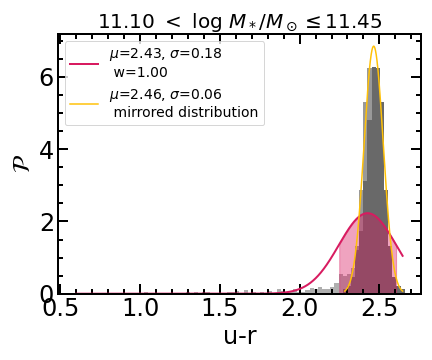}
    \caption{Examples of Gaussian-fit-based algorithms used to identify blue and red populations across different mass bins. The three cases illustrate the colour distribution of our simulated sample at $z \leq 0.08$ in three different mass bins, offering a clear demonstration of how our algorithm works (see the main text for full details). 
    The full colour distribution of galaxies in each bin is shown as a grey histogram and the Gaussian fits as curves, with the shaded regions indicating the standard deviation and the dashed line the intersection between the two curves (where present). The parameters for each Gaussian (mean, $\mu$, standard deviation, $\sigma$, and weight, $w$) are labelled in every panel. In the left and middle panels, our algorithm favours a double Gaussian distribution (blue and red, with the grey line indicating their sum). However, a real GV (green histogram) is identified only in the middle panel, since here the condition in Eq. \ref{eq:criterion_realGV} is met. In the right panel, the algorithm prefers a single Gaussian fit, which is shown as a red curve. In this case, a subsample of the total is extracted to obtain a symmetric distribution centred around the peak of the original. This sub-distribution, represented by the dark grey histogram, is fitted by the yellow Gaussian curve.  The parameters from this fit better represent the red population in this mass bin and are used in our analysis.}
    \label{fig:urdistribution_GMM}
\end{figure*}

\begin{figure*}[htb]
    \centering
    \includegraphics[width=0.8\columnwidth]{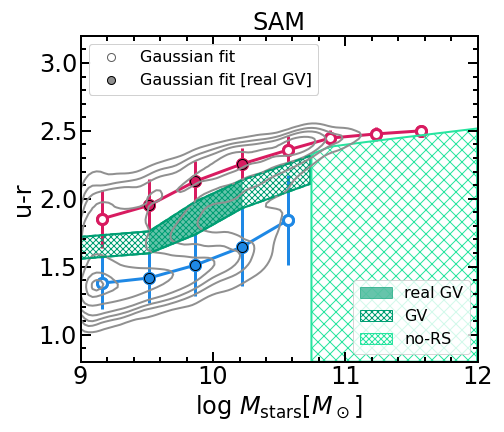}\quad
    \includegraphics[width=0.8\columnwidth]{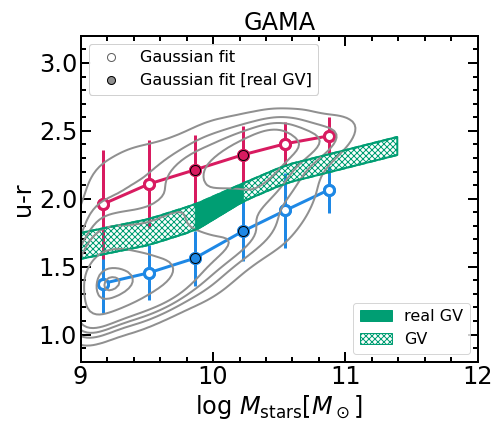}
    \caption{GV identification in the colour--mass diagram of both our simulated and observed samples at $z \leq 0.08$. The distribution of galaxies is shown as contours marking the  $1,\,5,\,20,\,50,\, 70,\,80,\,$ and $90\,\%$ percentiles of the distribution. The circles and error bars indicate the mean and standard deviation of the Gaussian fits performed by our algorithm in each mass bin for both the blue and red population. Filled circles are associated with bins in which the real GV can be identified (according to Eq. \ref{eq:criterion_realGV}). This region is depicted in solid green in the plots. The green hatched region represents an extension of this GV, obtained by using information from the Gaussian fits. The light green hatched region in the SAM sample marks the region of the diagram where  galaxies are identified as not belonging to the RS, since a single-Gaussian fit is favoured in these mass bins. In the GAMA sample, the GV has been linearly extrapolated in the high stellar mass range, where the number of galaxies per bin was not sufficient to perform a double-Gaussian fit ($<50$ galaxies).}
    \label{fig:GVdefinitionArlecchino}
\end{figure*}

The GV lacks a universally accepted definition, with its boundaries differing across studies and often being set by subjective empirical lines. Since our work involves defining the GV in both observed and simulated samples, we aim to establish a clear, objective definition that can be applied automatically, free from subjective criteria.

\subsection{Gaussian fits}
We adopted a Gaussian fitting approach to define the GV, which applies to any colour--mass distribution of galaxies. The sample was first divided into $N_{\rm bins}$ mass bins, with only bins containing more than $50$ objects considered. Within each bin, we used the Python class \texttt{GaussianMixture} to fit one and two Gaussians to the colour distribution. For the two-Gaussian fit, this class also provides the mixture weights of the resulting Gaussian components, indicating the significance of each component in the overall distribution. Additionally, it computes the  Bayesian information criterion (BIC) value, which helps balance the trade-off between the goodness of fit and the complexity of the model, favouring models that achieve a good fit with fewer parameters.

We applied three criteria to determine whether the distribution is better represented by one or two populations. 
\begin{itemize}
    \item[(i)] The BIC value must favour the two-Gaussian model. 

    \item[(ii)] None of the obtained weights, $w$, can be less than $w_{\rm min}=0.2$.

    \item[(ii)] The areas within one standard deviation in the two Gaussian curves must not overlap. 
\end{itemize}

\noindent If all conditions are met, we consider that the colour distribution is described by a two-Gaussian fit. The means $\mu$ and standard deviations $\sigma$ of these Gaussian components are used to help determine the presence of a GV in the distribution.
Specifically, denoting $x_{\rm BR}$ as the intersection point of the blue and red Gaussians, we required that 

\begin{equation}
\begin{cases}
\mu_{\rm R} - \sigma_{\rm R} \geq x_{\rm BR},\\
\mu_{\rm B} + \sigma_{\rm B} \leq x_{\rm BR},
\end{cases}
\label{eq:criterion_realGV}
\end{equation}
where the subscripts ${\rm B}$ and ${\rm R}$ refer to the blue and red Gaussians, respectively. This approach defines the GV, which we term the real GV, as the region of the colour distribution spanning from  $\mu_{\rm B} + \sigma_{\rm B}$ to $\mu_{\rm R} - \sigma_{\rm R}$.\\

The described conditions at work are explicitly shown in Fig. \ref{fig:urdistribution_GMM}, in the left and middle panels.
Our algorithm prefers a double-Gaussian fit for the distribution in both mass bins. However, in the $9 < {\rm log}\, M_*/M_\odot \leq 9.35$ mass bin, the two Gaussians do not meet the criterion specified in Eq. \ref{eq:criterion_realGV}, so a real GV is not identified. In contrast, the Gaussians in the $10.05 < {\rm log}, M_*/M_\odot \leq 10.40$ bin are well separated, allowing us to identify a real GV, represented by the green distribution in the histogram.\\

When the algorithm favours a single-Gaussian fit, we proceeded as follows. We first identified the peak of the entire distribution and isolated the portion extending from this peak to the nearest extreme. This segment was then mirrored to create a symmetric distribution centred on the original peak. We fitted this mirrored distribution with a single Gaussian, using its mean and standard deviation to represent the population.

An example of this process is illustrated in the right panel of Fig. \ref{fig:urdistribution_GMM}. The mirrored distribution, shown in dark grey, is fitted with a Gaussian (yellow curve), in contrast to the red curve used to fit the full distribution (light grey). This mirroring technique effectively removes the tail of the distribution, yielding Gaussian parameters that better represent the bulk of the objects in this region.\\

\begin{figure*}[h]
\centering
\includegraphics[width=\hsize]{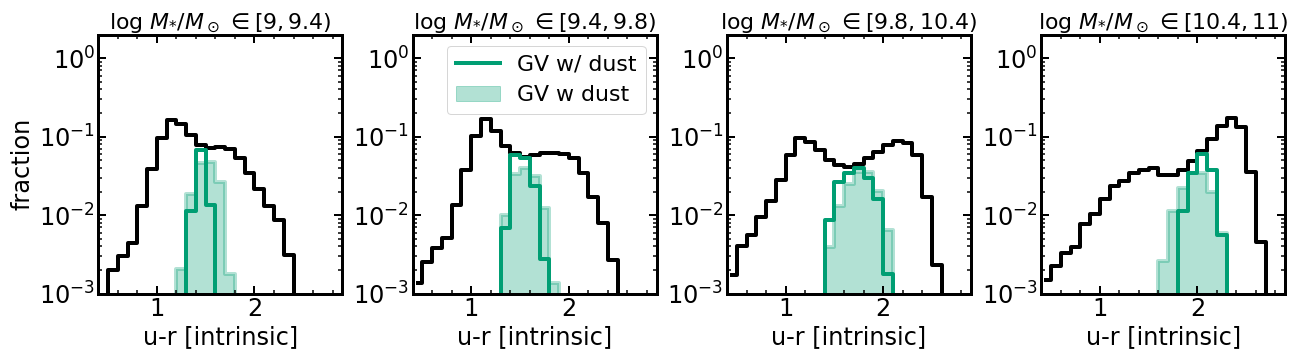}
  \caption{\rev{Intrinsic colour $(u-r)$ distribution in different stellar mass bins for our simulated galaxies at $z \leq 0.08$. Green lines and areas refer to GV galaxies as defined using, respectively, the intrinsic and dust-attenuated colour--mass diagram (Sect. \ref{sec:GVdefinition}).}}
  \label{fig:colorMass-intr}
\end{figure*}

\subsection{Building the GV}

The fitting procedure described above results in the Gaussian function parameters for each mass bin of the distribution. These parameters are then used to split the colour--mass diagram into different classes (particularly to identify a GV), as illustrated in Fig. \ref{fig:GVdefinitionArlecchino} for both the simulated and observed samples.

We began with the real GV, identified in the bins where the condition in Eq. \ref{eq:criterion_realGV} is satisfied. In these cases, the GV directly corresponds to the real GV, represented by the green shaded area in Fig. \ref{fig:GVdefinitionArlecchino}.

In the mass bins where the algorithm does not identify a real GV but still applies a two-Gaussian fit, we extended the real GV by using the Gaussian parameters and assigning a width based on that of the real GV. Specifically, we selected the Gaussian with the highest weight in the double-Gaussian fit. Given the mean and standard deviation of this Gaussian, the extended GV in this mass bin was defined with bounds of $\mu \pm \sigma$ and $\mu \pm (\sigma + \Delta^{\rm near}_{\rm real , GV})$, where $\Delta^{\rm near}_{\rm real , GV}$ represents the width of the real GV in the nearest mass bin where it has been identified. If two neighbouring mass bins have a real GV, the average width is used. The $\pm$ sign is chosen based on the Gaussian fit, with $+$ for blue and $-$ for red. This extended GV is shown as a hatched green region in Fig. \ref{fig:GVdefinitionArlecchino}. This approach allows us to define a GV in the colour--mass diagram for every mass bin where a two-Gaussian fit has been applied.\\

In cases where a single-Gaussian fit is used, we took a more conservative approach and chose not to define a GV. Instead, we divided the galaxy population into two categories, using $\mu \pm \sigma$ as the boundaries for these groups, with $-$ for a red peak and $+$ for a blue peak. If a red peak is identified, the categories are labelled RS and no-RS. If a blue peak is identified, they are labelled BC and no-BC. This scenario occurs in the most massive bins of our simulated sample, where only the red population is clearly defined, as indicated in light green in Fig. \ref{fig:GVdefinitionArlecchino}. Conversely, in high-redshift snapshots of our simulation, it is more common to find only a clear blue population in the less massive bins.\\

This definition allows us to identify a GV (where possible) for any given distribution in the colour--mass diagram. Once the GV is specified, the BC and RS are defined as everything below and above the GV, respectively (except in cases where no-BC or no-RS is identified). It is important to note that our approach is fairly conservative, as we avoided extrapolating the GV to mass bins where a clear colour bimodality was not observed. However, our findings remain largely unaffected by this definition. We verified that using alternative methods, such as extrapolating the GV where a single Gaussian is preferred or applying two simple linear fits to delimit the GV, would not significantly alter our conclusions.

All analyses presented below use the GV definition outlined above, specifically the region formed by the real GV and its extensions, where a two-Gaussian fit is preferred.

\section{The impact of dust attenuation}
\label{sec:dustatt}

It is well known that dust obscuration reddens the true, intrinsic colour of galaxies resulting from stellar emission. Consequently, some intrinsically blue, star-forming galaxies may appear redder (green or red in our classification). \rev{The impact of dust reddening is more important for star-forming galaxies \citep[e.g.][]{Schiminovic07} and sometimes leads to alternative approaches insensitive to dust obscuration  being adopted \citep[e.g. using the $4000\,\AA$ break;][]{Angthopo2019,Angthopo2020,Angthopo2021}. However, it has also been shown that the GV is a physical state, whose existence does not depend on the effects of dust attenuation on galaxy colours \citep{Wyder:2007,Salim2014}. Here we} exploited our simulated sample, for which both dust-attenuated and dust-free colours are available, to explore the impact of dust on shaping the colour-mass diagram.

\begin{figure*}[htb]
    \centering
    \includegraphics[width=0.8\columnwidth]{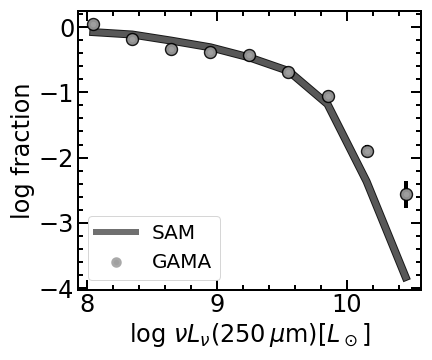}\quad
    \includegraphics[width=0.9\columnwidth]{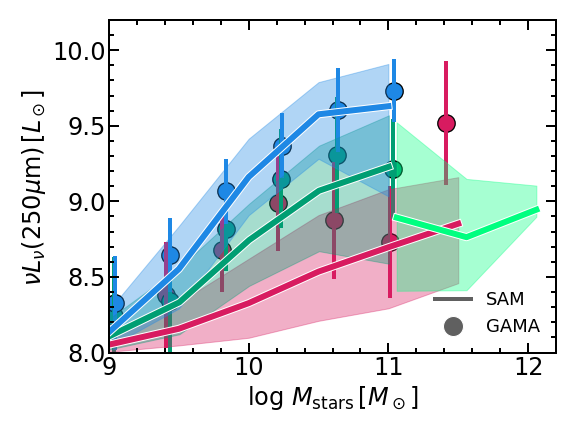}
    \caption{\rev{Comparison between the simulated (solid lines) and GAMA (filled circles) sample in terms of $250 \, \mum$ luminosity. Left panel: $\Lduecinquanta$ distribution of the two samples. Right panel: $M_{\rm stars}-\Lduecinquanta$ relation for the two samples split according to the classification outlined in Sect. \ref{sec:GVdefinition} (blue for BC, green for GV, red for RS, and light green for no-RS objects). Solid lines and filled circles are median values, while the error bars and shaded regions refer to the $16-84$th percentiles.}} 
    \label{fig:L250}
\end{figure*}

In our model, the BC and GV galaxies typically have more dust than equal-mass galaxies in the RS, which are mainly passive and gas-poor. As a result, dust reddening is more pronounced for galaxies in the BC and GV, where the intrinsic $u-r$ colour is reddened by $\sim 0.4-0.2$ dex. Attenuation is modest in RS galaxies ($\lesssim 0.2-0.1$ dex).

In light of this, when considering intrinsic colours the blue region of the colour--mass diagram shifts downwards by about $0.2-0.3$ dex, while intrinsic colours are nearly identical to dust-attenuated ones for red galaxies. The net effect of the attenuated-to-intrinsic transformation is a shift towards bluer $u-r$ colours, which is more pronounced for blue galaxies. Importantly, in this transformation, GV galaxies likely remain GV galaxies: although their colour becomes bluer, most dust-attenuated GV galaxies also have intrinsic green colours. In other words, the transition population identified in the dust-attenuated colour--mass diagram of Fig. \ref{fig:colorMass} mainly consists of galaxies with intrinsic intermediate colours. 
The percentage of significant outliers due to dust attenuation is very low. To give an idea, only $\sim 6.5\%$ of intrinsically blue galaxies move to the GV when dust attenuation is considered. As for the transformation of the GV, the $\sim 94\,\%$ of galaxies belonging to the intrinsic GV are within $0.2\,{\rm dex}$ of the dust-attenuated GV. These percentages were derived by defining BC, GV, and RS in the intrinsic colour--mass diagram using the same procedure detailed in Sect. \ref{sec:GVdefinition} for the dust-attenuated diagram.

\rev{Figure \ref{fig:colorMass-intr} provides further insight into the modest impact of dust on GV classification. It displays the intrinsic $u-r$ colour distribution of simulated galaxies in different stellar mass bins. Two definitions of the GV are adopted: one derived from the intrinsic colour--mass diagram (solid green lines) and another from the dust-attenuated diagram (as in the rest of the paper; green shaded areas). This comparison demonstrates that even when the GV is defined using a dust--attenuated diagram, the selected galaxies predominantly occupy the middle of the intrinsic colour distribution, overlapping with the GV population obtained from the intrinsic colour-mass diagram.}

We conclude that although dust obscuration plays a role in reddening galaxies, the bulk of the galaxies in a dust-attenuated GV belong to a transition region in the intrinsic colour--mass diagram.

\section{The sub-millimetre $250{\mu \rm m}$ emission}
\label{sec:250em}

In this section we analyse the sub-millimetre $250\,\mum$ emission of both simulated and observed $z\leq 0.08$ galaxies.
The simulated and observed $\Lduecinquanta$ distributions are shown in Fig. \ref{fig:L250} (left panel) down to \revv{${\rm log} \, \Lduecinquanta/L_\odot \sim 8$, where both the observed and simulated sample are complete.} 
\rev{We computed $L_\nu(250\mu\mathrm{m})$ luminosities for GAMA galaxies  
using k-corrections for a SED of a modified black body: $L_\nu \propto B_\nu(T) \ \nu^\beta$, where $B_\nu(T)$ is Planck's law for the spectral radiance of a black body of temperature $T$ and $\beta$ is the emissivity index. We followed \citet{Guo11} and assumed $T=28 K$ and $\nu=1.5$.}
In Fig. \ref{fig:L250} GAMA galaxies have been $1/V_{\rm max}$ weighted assuming a completeness flux of $S_{250}=27.8$ mJy \citep{Bourne:2016}.
\rev{The distributions are in fair good agreement, but in the high luminosity end, where the GAMA sample hosts a slightly larger number of luminous galaxies. }

\rev{The $250\, \mum$ luminosity as a function of stellar mass for the different classes defined in Sect. \ref{sec:GVdefinition} (BC, GV, RS, and no-RS) is shown in the right panel of Fig. \ref{fig:L250}. There is a general trend according to which $\Lduecinquanta$ grows from RS to BC galaxies at fixed stellar mass, for both our model and observations, and the agreement between the two samples is good. GV galaxies have $\Lduecinquanta$ lower than BC ones by a modest factor of $\sim 2$, while RS galaxies can feature $\Lduecinquanta$ values lower by up to an order of magnitude than the bluest galaxies. This is true up to ${\rm log }M_{\rm stars}/M_\odot \simeq 11$. Differences in $\Lduecinquanta$ between RS and no-RS galaxies disappear for larger mass galaxies in our simulated sample. This behaviour is due to the large fraction of rejuvenating galaxies in this mass range. These galaxies are those that were previously passive (i.e. they were in the RS), but now have bluer colours (meaning that they belong to the no-RS class, since a BC is not explicitly identified in this mass range) as a consequence of a rejuvenation event, for example a starburst. Despite featuring blue-to-green colours, these rejuvenating systems have relatively low $250 \, \mum$ luminosity (see the discussion in Sect. \ref{sec:rejuv}). Unfortunately, the low number of galaxies in the observed sample in the most massive bin is too low ($5$) to test this model prediction against observations.\\}

\rev{We conclude this section noting that \cite{Eales2018} observed that sub-millimetre detected galaxies (specifically those with $250 \, \mum$ emission $\geq 30 \,{\rm mJy}$ from $H-$ATLAS) are predominantly located in the optical GV. They suggested that this could originate from a continuous, non-bimodal distribution of galaxies in the $M_{\rm stars}-$SFR diagram, attributing the SFR bimodality to an incorrect mapping between colours and SFR (and vice versa). In other words, they interpret the observation of a `green mountain' as an argument against the widely accepted scenario that associates GV galaxies with a transition population (in terms of the SFR) .
Although this green mountain is in place when looking at the whole distribution (see their Fig. 3), it should be considered that GV (as well as BC and RS) colours evolve with stellar mass, likely reddening. Our analysis, which accounts for this, highlights that actually GV galaxies have intermediate $250 \, \mum$ emission.}

\begin{figure*}
\centering
\includegraphics[width=\hsize]{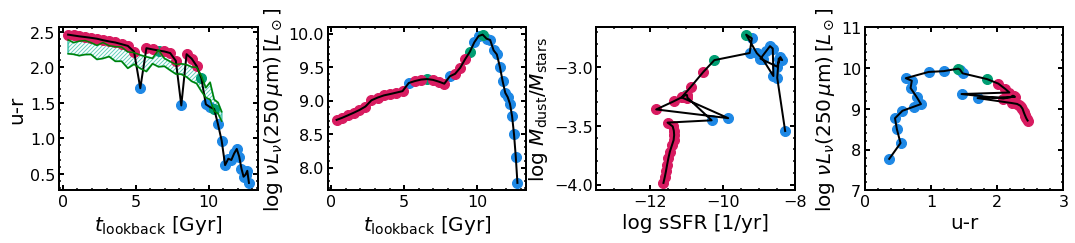}
  \caption{Evolution of a $z\simeq0$ RS galaxy with $M_{\rm stars}^{z=0} \simeq 10^{10.9}\,M_\odot$. \revv{Left panels: Time evolution of the rest-frame colour $(u-r)$ and $\Lduecinquanta$. Right panels : Evolution of the galaxy in the sSFR--$M_{\rm dust}/M_{\rm stars}$ and colour-$\Lduecinquanta$ diagrams.}
  Each point corresponds to a snapshot of the simulation and is coloured according to the galaxy position at that given time in the $M_{\rm stars}-$colour diagram (i.e. blue, green, and red for BC, GV, and RS, respectively) according to the $z$-dependent GV definition given \revv{in Sect. \ref{sec:250em_evo}}. The reference GV evolution is shown in the first panel.}
  \label{fig:red_evo}
\end{figure*}

\begin{figure*}
\centering
\includegraphics[width=\hsize]{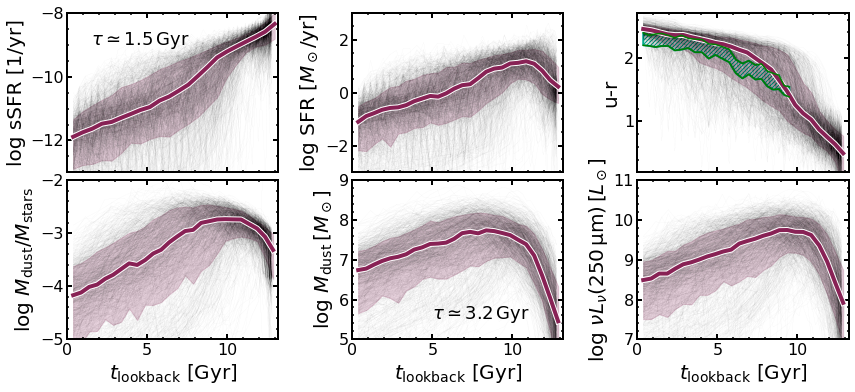}
  \caption{Time evolution of some key quantities for a sample of $z \sim 0$ red galaxies. In clockwise order, we show the sSFR, SFR, colour $u-r$, $\nu L_\nu ({\rm 250 \, \mu m})$, dust mass, and specific dust mass. An evolving GV (assuming the median mass evolution of the sample) is shown in the $(u-r)-{\rm time}$ plot. Tracks of individual galaxies in each diagram are shown as grey lines, and the median and $16-84$th percentiles dispersion of all the tracks in red. Finally, we report the typical $e$-folding timescales associated with the (decaying) evolution of the sSFR and dust mass.}
  \label{fig:MsGT10_evo}
\end{figure*}

\section{\rev{Evolution in the colour-$\Lduecinquanta$ diagram}}
\label{sec:250em_evo}

This section is devoted to carefully examining the sub-millimetre emission during the evolution of our model galaxies, with particular focus on the GV phase. It might be argued that the substantial sub-millimetre emission in GV galaxies occurs because these are fundamentally blue galaxies (i.e. star-forming galaxies) with a significant amount of dust, which reddens their optical colours and fuels the sub-millimetre emission. However, this is not the case, as already indicated in Sect. \ref{sec:dustatt} (Fig. \ref{fig:colorMass-intr}): only $\sim 6.5 \, \%$ of blue galaxies move to the GV as a result of dust attenuation. Model galaxies in the GV (on average) exhibit intermediate sSFR and intrinsic colour, suggesting that they represent a transitioning population rather than merely a dust-reddened one.

The reason for the sub-millimetre emission of GV galaxies is that, while these galaxies experience a rapid quenching that quickly reduces their sSFR and turns their colour green, the amount of dust in these objects remains relatively high, enabling a conspicuous $250 \,\mum$ emission. We show this comprehensively in the following sections.\\

\rev{Since we are interested in the evolution of the properties of our model galaxies, we needed to adopt an evolving definition of the GV, which is expected to vary with redshift. 
We used the definition of GV introduced in Sect. \ref{sec:GVdefinition} to identify it in all the snapshots of our simulations. The colour of the GV becomes progressively redder as redshift decreases, reflecting the colour evolution of the blue and red populations, which are both bluer at high redshift, as previously reported in observational and theoretical studies \citep[e.g.][]{Gu2018, Wright2019}. We only used $z \leq 2.25$ outputs, but this choice does not impact our results given the young age of the Universe at this redshift.}

\subsection{A red galaxy case study}
\label{sec:singlegalevo}

We began by analysing the evolution of a red galaxy at $z\simeq0$, with $M_{\rm stars}^{z=0} \simeq 10^{10.9}\,M_\odot$. This is illustrated in Fig. \ref{fig:red_evo}, which shows the evolution of the sSFR, the dust-to-stellar mass ratio ($M_{\rm dust}/M_{\rm stars}$), the colour $(u-r)$, and the $\duecinquanta$ emission. This particular galaxy was selected because its evolution is relatively smooth in terms of star formation, aside from a few bursty episodes, making it an ideal example for pedagogical purposes.
The galaxy is classified as part of the BC, GV or RS, represented by blue, green, and red points based on \rev{our} redshift-dependent definition of the GV.

The sSFR of the galaxy gradually decreases, leading to a reddening of its colour. Around $t_{\rm lookback}\simeq 10\,{\rm and}\,9 \, {\rm Gyr}$, the galaxy transitions into the GV and eventually the RS. The key point to emphasise is that during this transition from the BC to the RS, the galaxy colour reddens significantly, while $\Lduecinquanta$, though decreasing, remains comparable to its high levels during the BC phase. In other words, while the $u-r$ colour evolves rapidly and consistently towards redder values, the $\Lduecinquanta$ in the transition (GV) region stays relatively high. This results in a semi-circular trajectory in the colour-$\Lduecinquanta$ diagram, where the GV region is characterised by intermediate colours and high $\duecinquanta$ luminosities. This pattern is also reflected in other galaxy properties, particularly dust content and SFR, both normalised to stellar mass. Although the sSFR consistently decreases, the specific dust mass in the GV region remains quite high $(M_{\rm dust}/M_{\rm stars} \gtrsim 10^{-3})$, comparable to the values observed during the blue phase.

As the galaxy becomes redder, its $\Lduecinquanta$ continues to decline, along with its specific dust mass. During this phase, two noteworthy events occur at $t_{\rm lookback}\simeq 5\,{\rm and}\,8 \,{\rm Gyr}$, when the galaxy experiences two bursts of star formation. Each burst temporarily makes the galaxy bluer, moving it from the RS back to the BC for a single snapshot. We refer to galaxies undergoing these events as rejuvenated. However, during these rejuvenation-induced BC phases, the $\Lduecinquanta$ remains quite low because the dust production has not had sufficient time to accumulate enough mass to produce significant $\duecinquanta$ emission. We discuss some interesting implications of this in Sect. \ref{sec:rejuv} concerning rejuvenating systems. \\

\subsection{A red statistical sample}

\begin{figure}
\centering
\includegraphics[width=0.9\hsize]{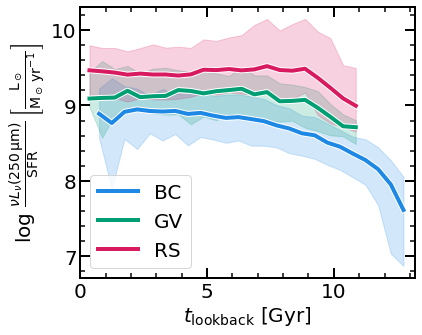}
  \caption{Evolution of $\nu L_\nu (250 \mum)/$SFR for the sample of \revv{$z\sim 0$} RS galaxies reported in Fig. \ref{fig:MsGT10_evo}, distinguishing their different stages during the evolution (BC, GV, and RS, respectively in blue, green, and red). Solid lines and shaded areas refer to the median trends and $16-84$th percentiles dispersions.}
  \label{fig:S250SFR_RGB}
\end{figure}

We now turn to the analysis of the evolution of the same key quantities in a large statistical sample of $\sim 2,000$ red galaxies at $z \simeq 0$, namely, the sSFR, dust mass, $(u-r)$ colour, and $\Lduecinquanta$, and are shown in Fig. \ref{fig:MsGT10_evo}. This analysis, averaged over the entire sample, highlights the key features relevant for interpreting the sub-millimetre flux during the GV transition.

Both the sSFR and $(u-r)$ colour evolve monotonically towards a passive (or red) state. However, this is not the case for the dust content (and specific dust content), which reaches a maximum at $t_{\rm lookback} \sim 9 \, {\rm Gyr}$. Notably, this peak does not coincide with the peak in the SFR, which occurs earlier at $t_{\rm lookback} \sim 11 \, {\rm Gyr}$. This discrepancy is due to two main factors. First, once dust is produced by stars, it remains in the ISM until it is removed by other processes such as astration, galactic outflows, or destruction by supernovae. On the contrary, the SFR is an instantaneous quantity, meaning that it keeps track of a relatively short period of the evolution of the galaxy. Second, and more importantly, the dust budget can increase due to grain growth in the ISM, leading to a delay of $\gtrsim 1 \, {\rm Gyr}$ between the SFR peak and the dust abundance peak.

Another way to examine this is by comparing the $e$-folding decay timescales of the sSFR and dust abundance, as shown in Fig. \ref{fig:MsGT10_evo}. The timescale for dust abundance is roughly twice as long as that for sSFR, indicating a slower decline in dust content. \rev{This estimate is obtained by averaging over the whole population. When looking at the $e$-folding decay timescales for individual galaxies, the ratio $\tau_{\rm dust}/\tau_{\rm sSFR}$ is still typically larger than $1$; it is $\simeq 1.3$ for log $M_{\rm stars}/M_\odot \lesssim 10.5$ galaxies and $\gtrsim 2$ for log $M_{\rm stars}/M_\odot \gtrsim 11$ galaxies.}\\

An interesting consequence of the time lag between dust emission and star formation activity is that sub-millimetre emission traces the SFR differently in BC, GV, and RS galaxies. This is explicitly illustrated in Fig. \ref{fig:S250SFR_RGB}, which depicts the time evolution of the ratio between $\Lduecinquanta$ and SFR for the sample of galaxies discussed in this section, while distinguishing between the different evolutionary stages (BC, GV, and RS). As galaxies move from BC to RS, this ratio increases, reflecting a sharper decline in SFR compared to the decrease in sub-millimetre luminosity (and dust mass). This analysis highlights the limitations of using sub-millimetre fluxes as direct tracers of SFR. \revv{We also note that BC galaxies at high redshifts feature $\Lduecinquanta/{\rm SFR}$ values lower by up to three orders of magnitude than the local galaxy population, as a consequence of generally lower dust-to-SFR ratios (and higher sSFR) in early galaxies.}

\subsection{Rejuvenating systems}

\label{sec:rejuv}

\begin{figure*}
\centering
\includegraphics[width=0.9\hsize]{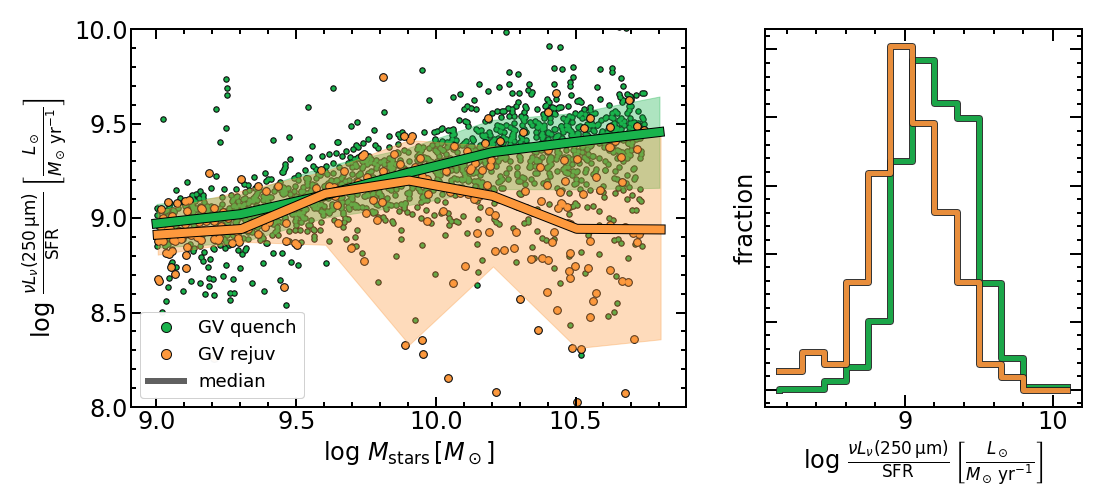}
  \caption{\rev{Left panel: $\Lduecinquanta$ normalised to the SFR as a function of stellar mass for $z \simeq 0$ GV galaxies (represented by filled circles). Quenching GV galaxies are in green and rejuvenating systems (transitioning from the RS to the GV) in orange. Median trends of the two galaxy samples are shown as solid lines, while shaded areas refer to the $16-84$th percentile dispersions.
  Right panel: Distribution of the $\Lduecinquanta$ normalised to the SFR for both quenching and rejuvenating GV galaxies (green and orange, respectively).}}
  \label{fig:rejuv}
\end{figure*}

The differential variation in colour and sub-millimetre flux has important implications for rejuvenating systems within the GV. Not all galaxies in the GV come directly from the BC. Some of them are undergoing a rejuvenation process, marked by a reverse quenching scenario in which galaxies shift from a quiescent to a star-forming state, often driven by factors such as increased gas supply, which is frequently the result of mergers in our SAM.

\rev{In Fig. \ref{fig:rejuv} (left panel), we show the predicted $\Lduecinquanta$ normalised to SFR as a function of stellar mass for $z \simeq 0$ galaxies in the GV. We differentiate between quenching GV galaxies and rejuvenating systems transitioning back from the RS to the GV. Rejuvenating galaxies are operationally defined as those that spent the two last snapshots in the RS just before moving to the GV. While this definition can be somewhat ambiguous due to the episodic nature of star formation and colour evolution, it helps identify galaxies that are not simply transitioning from a star-forming state. However, we also verified that alternative plausible definitions\footnote{These include considering galaxies that, just before moving to the GV, have spent \textit{(i)} at least $5$ out of $10$ snapshots in the RS, and \textit{(ii)} 3 consecutive snapshots in the RS.} of rejuvenating galaxies yield similar results.} Typically, rejuvenating systems have $\Lduecinquanta/$SFR values lower than the GV average. As previously discussed in Sect. \ref{sec:singlegalevo}, this occurs because rejuvenating systems experience bursts of star formation that temporarily make their colours bluer, but they maintain relatively low dust masses, leading to lower $\duecinquanta$ emission.

\rev{It is important to note that this result is not just a consequence of any different levels of SFR during the quenching and rejuvenating paths. Actually, at fixed stellar mass, SFRs for rejuvenating galaxies are lower than those in quenching galaxies by a modest factor of $\sim 1.6$, while dust masses are lower by up to a factor of $\sim 2.5$.}\\

In summary, sub-millimetre emission is a valuable indicator for identifying rejuvenating systems within the GV. Although these systems have green colours similar to other GV galaxies, their sub-millimetre flux is on average lower, due to their shorter time available to rebuild their dust content.\\

\subsection{The colour-$\Lduecinquanta$ diagram}

\begin{figure}
\centering
\includegraphics[width=0.9\hsize]{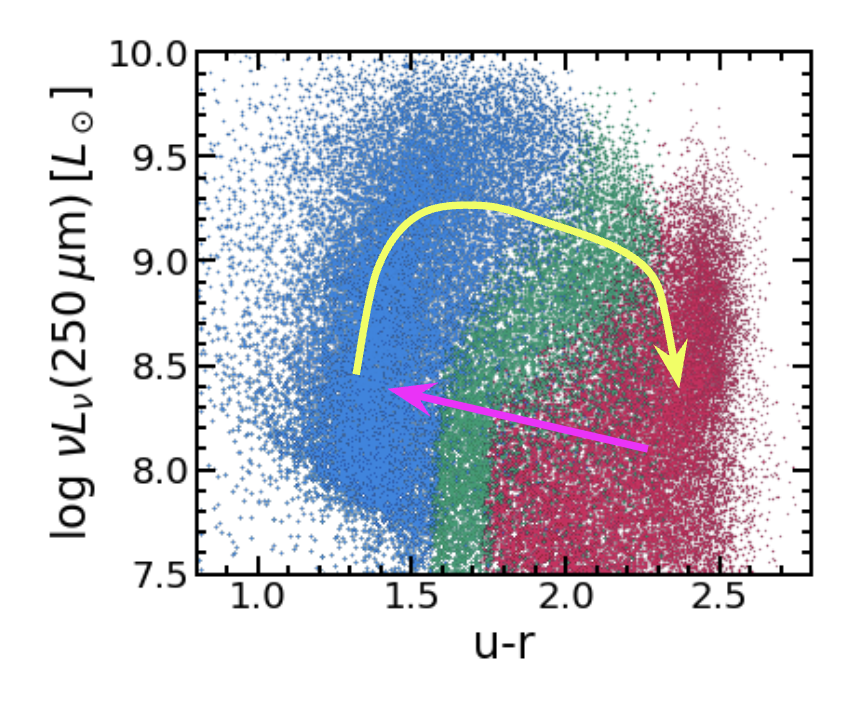}
  \caption{\rev{Colour-$\Lduecinquanta$ diagram for our $z \leq 0.08$ model galaxies in the BC, GV, and RS (represented by blue, green, and red points, respectively). The yellow and magenta arrows schematically represent, respectively, the quenching and rejuvenating paths of the galaxies in this diagram.}}
  \label{fig:color_submm_evo}
\end{figure}

To conclude \rev{and summarise our results}, we show the colour-$\Lduecinquanta$ diagram for our model galaxies at $z \leq 0.08$ in Fig. \ref{fig:color_submm_evo}. \rev{Building on the previous analysis of evolutionary trends, we depict a schematic quenching path in this diagram using a yellow arrow, while the magenta arrow indicates a rejuvenating path. The key message is that sub-millimetre emission increases during the BC phase, remains high throughout the GV quenching transition, and then decreases once the galaxy evolves into the RS. Galaxies that undergo rejuvenation will transit this diagram from red to blue colours, but still featuring low sub-millimetre emission.}

\section{Discussion and conclusions}

In this work we studied the $250\, \mum$ sub-millimetre emission of galaxies during their transition through the GV. We utilised predictions from a SAM of galaxy evolution that incorporates processes for the formation and evolution of dust grains within a galactic context. These predictions were employed to estimate observable properties of galaxies, including sub-millimetre emission, using the RT code GRASIL, which simulates both dust attenuation and the thermal emission of dust grains.\\

We concentrated on galaxies at $z \leq 0.08$, whose distribution in the $u-r$ colour--mass diagram (Fig. \ref{fig:colorMass}) aligns well with GAMA observations.\ We see a clear colour bimodality, although there are some tensions, especially at low and high stellar masses. To analyse the properties of galaxies in the GV, we introduced a new, precise GV definition that can be applied to any colour--mass diagram distribution (Sect. \ref{sec:GVdefinition}). This allowed us to consistently compare our simulation results and observational data, although there is no perfect match in terms of absolute colours.

With our new definition, we analysed various properties of the different galaxy populations identified in the colour--mass diagram, mainly the BC, GV, and RS. We find that dust-induced attenuation does not significantly alter the distribution of our model galaxies in this diagram (Fig. \ref{fig:colorMass-intr}). Galaxies in the GV, as seen in the dust-attenuated colour--mass diagram, are generally also identified as transitioning galaxies in a dust-free (or intrinsic) colour--mass diagram.

When examining sub-millimetre emission across different populations, \rev{we find that, at fixed stellar mass, GV galaxies exhibit $\duecinquanta$ luminosities that are approximately half those of BC galaxies, while RS galaxies have luminosities lower by up to an order of magnitude  than BC galaxies. This is true for our simulated galaxy sample and GAMA observations (Fig. \ref{fig:L250}). These differences are present for stellar masses of log $M_{\rm stars}/M_\odot \lesssim 11$. At higher stellar masses, our model predicts similar $\Lduecinquanta$ \revv{values for RS and no-RS galaxies. The distinction between RS and no-RS galaxies is necessary, as the BC and GV are not distinctly identifiable at these masses}. We interpret this finding in the context of the numerous rejuvenating systems outside the RS in this mass range, which have blue or green colours but relatively low sub-millimetre emission.}

\begin{figure}
\centering
\includegraphics[width=0.85\hsize]{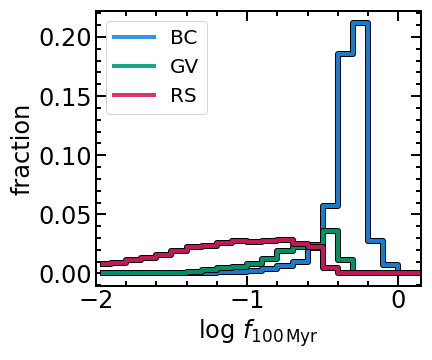}
  \caption{Distribution of the fraction of \revv{bolometric stellar} luminosity due to star formation, here evaluated as a stellar population younger than $100\,{\rm Myr}$ for BC, GV, and RS galaxies at $z = 0.03$.}
  \label{fig:youngcontr}
\end{figure}

We used our model to further investigate the sub-millimetre emission of galaxies during the GV phase. The sSFR of galaxies evolves (on average) in a generally monotonic fashion, causing their colour to gradually redden. In contrast, $\duecinquanta$ emission -- which is associated with the dust content in galaxies -- increases during the BC phase and remains high as galaxies transition towards the RS (Figs. \ref{fig:red_evo} and \ref{fig:MsGT10_evo}).

In addition to the dust $250\,\mum$ emission being different as galaxies evolve from the BC to the RS, the heating sources are also different. While $250\,\mum$ emission in BC galaxies is mainly powered by star-forming regions, as the galaxies transit to the RS, the importance of the ambient radiation from old populations becomes more important. This is shown in Fig. \ref{fig:youngcontr}, where the distribution of the fraction of \revv{the bolometric stellar} luminosity of star-forming radiation is shown for the BC, GV, and RS categories. The star-forming radiation is evaluated as the luminosity of stellar populations younger than $100\,{\rm Myr}$.

Finally, we have highlighted that sub-millimetre emission can help us identify rejuvenating GV galaxies. Specifically, at a given stellar mass, GV galaxies undergoing a rejuvenation event exhibit median $\duecinquanta$ emission that is $0.1-0.4$ dex lower compared to \rev{quenching} GV galaxies (Fig. \ref{fig:rejuv}). This lower emission is due to rejuvenating galaxies becoming bluer due to a starburst, not having had sufficient time to accumulate enough dust to produce significant sub-millimetre emission. Testing this prediction on an observed sample, along with other techniques for identifying rejuvenating systems, would be extremely interesting.
To refine our understanding of the nature of transition galaxies, in subsequent works we will extend our multi-wavelength analyses by incorporating radio data; this could offer a more complete view of the interplay between dust buildup, SFHs, and sub-millimetre/radio emission in these systems \citep[e.g.][]{Bressan2002}.\\

In conclusion, we remark on the importance of the galaxy evolution model that we adopted.\ It self-consistently tracks dust formation and evolution and formed the basis of our analysis. The model naturally predicts the delay between star formation and dust content evolution, along with the associated changes in optical colours and sub-millimetre emission. \rev{Interestingly, a faster decline in the SFR with respect to dust content has also been found in a series of observational work studying the ISM of early-type galaxies \citep{Micha19,Lesnieska23,Micha24,Nadolny24}.} Moreover, \rev{our model} shows that a high (low) sub-millimetre emission is not always a failsafe signature of high (low) star formation activity. GV galaxies in the quenching process can still feature sub-millimetre emission similar to those of BC galaxies, while rejuvenating systems, despite being significantly star-forming, can have relatively low sub-millimetre emission.

\label{sec:conclusions}

\begin{acknowledgements}
\rev{We thank the anonymous referee for the detailed report and useful suggestions which substantially improved this work.} This project has received funding from the Consejo Nacional de Investigaciones Cient\'ificas y T\'ecnicas (CONICET) (PIP-2021-11220200102832CO, PIP-2022-11220210100064CO), from the Agencia Nacional de Promoción de la Investigación, el Desarrollo Tecnológico y la Innovación de la Rep\'ublica Argentina (PICT-2018-3743, PICT-2020-3690), from the Secretaría de Ciencia y Tecnología de la Universidad Nacional de Córdoba, and from the European Union's HORIZON-MSCA-2021-SE-01 Research and Innovation Programme under the Marie Sklodowska-Curie grant agreement number 101086388 - Project LACEGAL. 
\end{acknowledgements}

\bibliographystyle{aa} 
\bibliography{GV} 

\end{document}